\documentclass[aps,prl,twocolumn,superscriptaddress,showpacs]{revtex4}

\usepackage[dvipdfmx]{graphicx, color}
\begin{document}


\title{Large entropy change derived from the orbitally assisted three-centered two-electron $\sigma$ bond formation in a metallic Li$_{0.33}$VS$_{2}$}

\author{N. Katayama}\email{katayama@mcr.nuap.nagoya-u.ac.jp}
\affiliation{Department of Applied Physics, Nagoya University, Nagoya 464-8603, Japan}
\author{S. Tamura}
\affiliation{Department of Applied Physics, Nagoya University, Nagoya 464-8603, Japan}
\author{T. Yamaguchi}
\affiliation{Department of Physics, Chiba University, Chiba 263-8522, Japan}
\author{K. Sugimoto}
\affiliation{Center for Frontier Science, Chiba University, Chiba 263-8522, Japan}
\author{K. Iida}
\affiliation{Neutron Science and Technology Center, Comprehensive Research Organization for Science and Society (CROSS), Tokai, Ibaraki 319-1106, Japan}
\author{T. Matsukawa}
\affiliation{Frontier Research Center for Applied Atomic Sciences, Ibaraki University, Tokai, Ibaraki 319-1106, Japan}
\author{A. Hoshikawa}
\affiliation{Frontier Research Center for Applied Atomic Sciences, Ibaraki University, Tokai, Ibaraki 319-1106, Japan}
\author{T. Ishigaki}
\affiliation{Frontier Research Center for Applied Atomic Sciences, Ibaraki University, Tokai, Ibaraki 319-1106, Japan}
\author{S. Kobayashi}
\affiliation{Department of Applied Physics, Nagoya University, Nagoya 464-8603, Japan}
\author{Y. Ohta}
\affiliation{Department of Physics, Chiba University, Chiba 263-8522, Japan}
\author{H. Sawa}					
\affiliation{Department of Applied Physics, Nagoya University, Nagoya 464-8603, Japan}
\date{\today}

\begin{abstract}
We discuss herein the emergence of a large entropy change in metallic Li$_{0.33}$VS$_2$ derived from the orbitally assisted loose $\sigma$ bond formation. Comprehensive structural studies based on synchrotron x-ray and neutron diffraction analyses clarify the fabrication of ribbon chains at 375 K, consisting of multiple three-centered two-electron $\sigma$ bonds based on the viewpoint of local chemical bonding. Although the metallic conductivity persists down to the lowest temperature measured, exceptionally large entropy change as a metal, as much as $\Delta$$S$ = 6.6 Jmol$^{-1}$K$^{-1}$, appears at the transition. Emergence of a large entropy change in a metallic state expects us the possible novel functional materials, such as a heat-storage material with rapid thermal response.

\end{abstract}

\maketitle

The search for various self-organizing phenomena in which multiple degrees of freedom of electrons intertwine in solids has been intensively studied. Examples include stripe-like charge ordering in copper and nickel oxides \cite{rf:1}, and the spontaneous formation of various molecular clusters in low-dimensional and/or geometrically frustrated systems \cite{rf:2,rf:3,rf:4,rf:5,rf:6,rf:7,rf:8,rf:9}. The driving force of the latter example is the formation of molecular orbitals accompanied by periodic modulation of the lattice structure, which gives rise to strong metal-metal bonds between the adjacent transition metal ions. 

The variety of molecular cluster patterns is strongly related to the number of $d$ electrons. In $d^1$-electron systems, dimers usually form between adjacent transition metal ions; MgTi$_2$O$_4$ with a spinel lattice is an example \cite{rf:3}. For $d^2$-electron systems, triangular-shaped trimers appear, as demonstrated by LiVO$_2$ and LiVS$_2$, which have two-dimensional triangular lattices \cite{rf:4,rf:4-1,rf:4-2}. When a noninteger number of $d$ electrons are involved in the bonds, complicated cluster patterns appear, such as for the octamer in CuIr$_2$S$_4$ \cite{rf:6}. Although the cluster patterns vary, the mechanism of the cluster formation can be generally understood based on the concept of orbitally induced Peierls state, proposed by Khomskii and Mizokawa \cite{rf:11}.  From the viewpoint of the local bonding, the resulting clusters are consisting of multiple dimers with two-centered two-electron $\sigma$ bonds (2c-2e) \cite{rf:2,rf:3,rf:4,rf:5,rf:6,rf:7,rf:8,rf:9}.

For a two-dimensional triangular lattice system with a $d$-electron number of 4/3, ribbon chain one-dimensional clusters consisting of multiple linear trimers appear \cite{rf:10}. The examples include $M$Te$_2$ ($M$ = V, Nb, Ta), as we will describe the detail later \cite{rf:19,rf:19-2}. Whangbo $et~al.$ proposed the ribbon chain formation can be understood based on the concepts of both hidden one-dimensional Fermi-surface nesting in $d$ orbitals connected to form a one-dimensional chain and local chemical bonding \cite{rf:10}. Although the former is basically consistent with an orbitally induced Peierls state introduced above \cite{rf:11}, the resulting local bonds constituting of linear trimers can be interpreted as a loose three-centered two-electron $\sigma$ bond (3c-2e), which sharply differs from the tight 2c-2e bond \cite{rf:2,rf:3,rf:4,rf:5,rf:6,rf:7,rf:8,rf:9}. Considering that various physical properties and functions derived from bonding are central themes in condensed matter physics \cite{rf:tamura,rf:112}, such as controlling of magnetism \cite{rf:collapsed}, anomalous metallic state \cite{rf:4,rf:9,rf:Li2RuO3} and bond breaking superconductivity \cite{rf:7,rf:nohara}, experimental investigations of loose 3c-2e bonds should extend possible methodologies, leading to chances to obtain novel functional materials.

In this paper, we present the crystal structure and properties of the layered transition-metal dichalcogenide Li$_{0.33}$VS$_2$, which, upon cooling, undergoes a trigonal-to-monoclinic phase transition accompanied by ribbon chain formation consisting of multiple 3c-2e bonds. We show that the entropy of Li$_{0.33}$VS$_{2}$ changes by as much as $\Delta$$S$ = 6.6 Jmol$^{-1}$K$^{-1}$ at the transition, although it has high electrical conductivity down to the lowest temperature measured. We discuss the exceptionally large entropy change in a metallic state is derived from the orbitally assisted loose 3c-2e bond formation, possibly leading to the novel functional materials, such as heat-storage materials with rapid thermal response.


Powder samples of Li$_{0.33}$VS$_{2}$ were prepared using a soft-chemical method followed by a solid-state reaction. Initially, Li-deﬁcient Li$_{\sim}$$_{0.75}$VS$_{2}$ was obtained by reacting an appropriate amount of Li$_{2}$S, V and S in an evacuated quartz tube at 700 $^\circ$C for three days. The products were immersed in a 0.2 M $n$-BuLi hexane solution for two days to obtain LiVS$_2$. Next, the Li content was quantitatively tuned to obtain Li$_{0.33}$VS$_2$ by using I$_2$ acetonitrile solution by using the equation, LiVS$_2$+ 0.33I$_2$  →Li$_{0.33}$VS$_{2}$+0.67LiI. The samples were characterized by both neutron and synchrotron powder x-ray diffraction experiments. Neutron diffraction experiment was done by using $\sim$10g of powdered samples in iMATERIA beamline equipped at J-PARC, Japan \cite{rf:iMATERIA}. For analysis, the Z-Rietveld software \cite{rf:Z-Rietveld1, rf:Z-Rietveld2} was used. Synchrotron powder x-ray diffraction experiments were done at BL5S2 beamline at Aichi Synchrotron, Japan. RIETAN-FP software \cite{rf:Rietan} was employed for Rietveld analysis. Differential scanning calorimetry (DSC) was conducted by using DSC 204 F1 Phoenix (Netzsch). Magnetic susceptibility was measured by a SQUID magnetometer (Quantum Design). Electrical resistivity was measured using the four-probe method. First principles calculation was performed using the WIEN2k code \cite{rf:Wien2k}. The computational details are available in the Supplementary Materials \cite{rf:Supple1}. 

Figures \ref{fig:Fig1}(a)-\ref{fig:Fig1}(c) show the temperature dependence of the magnetic susceptibility, electrical resistivity and DSC signals, respectively, for Li$_{0.33}$VS$_2$. This material is reported to undergo a first-order transition with a decrease in magnetic susceptibility at 375 K \cite{rf:18}. The present data show a similar temperature dependence, as shown in Fig. \ref{fig:Fig1}(a). The results of the neutron diffraction experiment show no sign of magnetic ordering at 200 K, as shown in the inset of Fig. \ref{fig:Fig1}(a). The magnetic susceptibility increases slightly at the lowest temperature, which may be attributed to paramagnetic impurities at concentrations below 1\%. To measure electrical resistivity, we used a low temperature sintered body, because the interlayer Li ions are easily defected when the sample is sintered at high temperature. Increasing the temperature through the phase transition at 375 K leads to a jump in the electrical resistivity; the low temperature phase maintains metallic conductivity of several m$\Omega$cm. Considering that the low temperature sintered sample is used, the intrinsic electrical resistivity is empirically one order of magnitude lower, indicating that Li$_{0.33}$VS$_2$ is metallic over the entire temperature range. Therefore, both phases above and below 375 K are Pauli paramagnetic phases, so the jump in magnetic susceptibility at the transition should be attributed to the difference in the electronic density of states. The electrical resistivity measurements also show anomalies at 294 K, which appear in all measured samples. However, the origin of the anomaly is unclear because no anomalies appear at the corresponding temperature in other experiments, such as magnetic susceptibility, DSC, and structural analysis, maybe indicating the anomaly is derived from the surface effect.

\begin{figure}
\includegraphics[width=8.2cm]{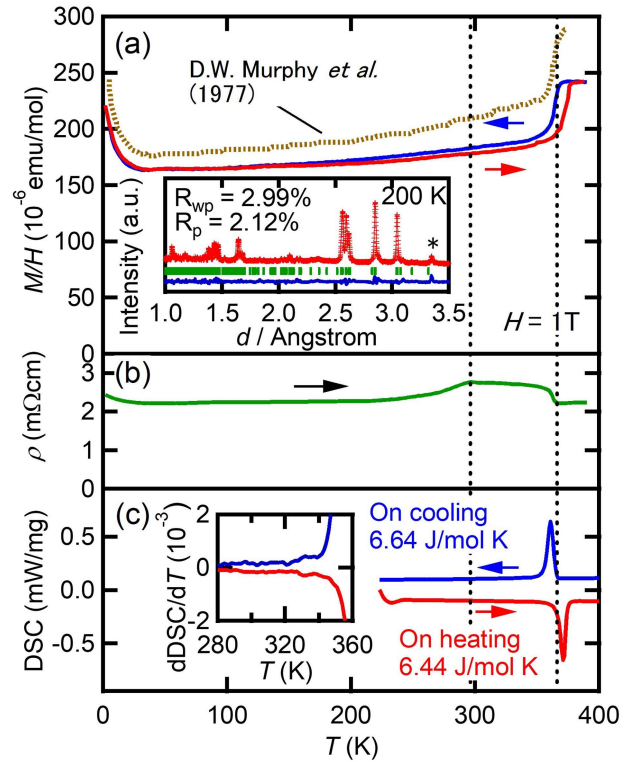}
\caption{\label{fig:Fig1}
(Color online) (a) Magnetic susceptibility of Li$_{0.33}$VS$_{2}$. Magnetic susceptibility data reported by D.W. Murphy $et~al.$ are taken from Ref.\cite{rf:18}. The inset shows the Rietveld refinement of neutron diffraction data at 200 K. Although a small impurity peak appears (see the asterisk), magnetic Bragg peaks are absent. (b) Electrical resistivity of Li$_{0.33}$VS$_2$. Because a large change in volume occurs at the transition at 375 K, cracks are generated in the sintered body, and the electrical resistivity as a function of temperature upon cooling does not match that upon heating. Only the cooling process is shown in (b). (c) DSC of Li$_{0.33}$VS$_2$. The inset shows the differentiated DSC data. No anomaly appears around 294 K. }
\end{figure}

As shown in Fig. \ref{fig:Fig1}(c), the DSC experiment clarifies the huge change in entropy of $\Delta$$S$ = 6.6 Jmol$^{-1}$K$^{-1}$ accompanied by the first-order phase transition at 375 K. While the large entropy change often appears in localized electron systems, such as in LiVS$_2$ with $\Delta$$S$ = 6.4 Jmol$^{-1}$K$^{-1}$ and in LiVO$_2$ with $\Delta$$S$ = 14.5 Jmol$^{-1}$K$^{-1}$ \cite{rf:4}, the entropy change of $\Delta$$S$ = 6.6 Jmol$^{-1}$K$^{-1}$ is extremely large for a metal to metal electronic phase transition, indicating the unusual feature of the transition in Li$_{0.33}$VS$_2$. Assuming that Li$_{0.33}$VS$_2$ is a metal weakly correlated with Wilson ratio $R_W$ $\sim$ 1 in the whole temperature range, which gives an estimate of the maximum entropy change as opposed to assuming other values for $R_W$, we can roughly estimate the entropy change to be $\Delta$$S$ = $\Delta$$\gamma$$T_c$ $\sim$ 1.5 Jmol$^{-1}$K$^{-1}$. It is extremely small compared with the experimental value of $\Delta$$S$ = 6.6 Jmol$^{-1}$K$^{-1}$, which seems to indicate that the transition involves ordering of degrees of freedom, such as orbital. 

\begin{figure}
\includegraphics[width=8.2cm]{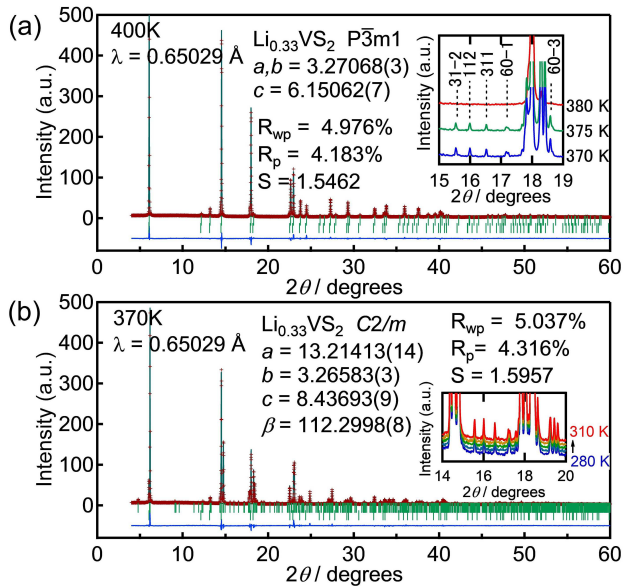}
\caption{\label{fig:Fig2}
(Color online) Synchrotron x-ray diffraction data at (a) 400 K and (b) 370 K. The inset of (a) shows the spectrum around 375 K. The inset of (b) shows the spectrum around 294 K. No significant change appears corresponding to the anomaly observed in the resistivity measurements.}
\end{figure}

To clarify the ground state of the low-temperature phase based on the structural analysis, we made comprehensive neutron and synchrotron x-ray diffraction studies. The neutron diffraction experiment indicates a Li fraction of 0.3388(13), which is very close to 1/3. The neutron diffraction experiment also clarifies that magnetic Bragg peaks do not emerge at the 375 K transition, as discussed before. Analysis of synchrotron x-ray diffraction data gives the structural parameters at each temperature. As shown in Fig. \ref{fig:Fig2}(a), the high-temperature phase is refined as $P$$\bar{3}$$m$1 with a regular V triangular lattice. Although $\sim$4.5\% of the Li$_{0.5}$VS$_2$ impurity is identified, it is successfully co-refined with Li$_{0.33}$VS$_2$. As shown in the inset of Fig. \ref{fig:Fig2}(a), the spectrum drastically changes below the transition temperature, and the low-temperature phase is refined by using the monoclinic space group $C$2/$m$, as shown in Fig. \ref{fig:Fig2}(b). Table SI in the supplementary material summarizes the structural parameters \cite{rf:Supple2}.

\begin{figure}
\includegraphics[width=8.2cm]{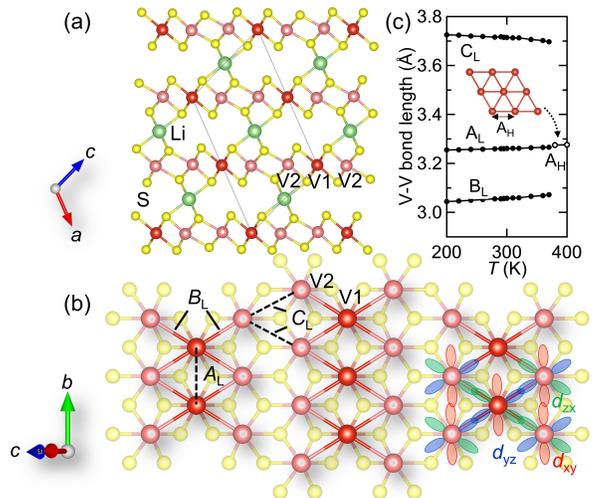}
\caption{\label{fig:Fig3}
(Color online) (a), (b) Refined crystal structure of Li$_{0.33}$VS$_2$. A$_L$, B$_L$ and C$_L$ indicate the V1-V1, V1-V2 and interchain V2-V2 bond lengths, respectively. Ribbon chain are clearly apparent in (b). (c) V-V bond length as a function of temperature. }
\end{figure}

Figures \ref{fig:Fig3}(a) and \ref{fig:Fig3}(b) show the low-temperature crystal structure obtained from the Rietveld analysis. At 375 K, Li ordering appears to form a one-dimensional chain in the direction of $b$ axis. In the low-temperature phase, the V site splits into two sites, V1 and V2 in the ratio of 1 : 2. As shown in Fig. \ref{fig:Fig3}(b), the V1 sites form a one-dimensional chain in the direction of $b$-axis, whereas V2 approaches the one-dimensional V1 chain. The result is that the ribbon chain in the low-temperature phase is clearly apparent. Fig. \ref{fig:Fig3}(c) shows the temperature dependence of V-V distances. The distance A$_L$ between adjacent V1 sites remains essentially constant across the transition, whereas the V1-V2 distance B$_L$ within the ribbon chain decreases to 3.075 \AA~at 360 K, which is about 8\% shorter than the V-V distance A$_H$ in the high-temperature phase. Relatively, the V2-V2 distance C$_L$ is significantly increased.

%

The mechanism at work in the phase transition may be understood by invoking Whangbo's theory \cite{rf:10}. In the high-temperature phase, the ${xy}$, ${yz}$, and ${zx}$ bands constructed by the corresponding $d$ orbitals are 4/9 filled. The $yz$ and $zx$ one-dimensional chains undergo a Peierls transition accompanied by a trigonal-to-monoclinic transition, which also broadens the $yz$ and $zx$ bands. As a result, the filling of $yz$ and $zx$ bands increases up to 2/3, forming linear trimers consisting of 3c-2e bonds. In contrast, $xy$ band becomes almost empty. The change in band filling, aided by Li-ion ordering, may be responsible for a large change in entropy of $\Delta$$S$ = 6.6 Jmol$^{-1}$K$^{-1}$. Although the contribution of the Li-ion order to the change in entropy is not clear, we consider it to be negligibly small based on the previous studies of the surrounding materials \cite{rf:Ag, rf:Levy}.

\begin{figure}
\includegraphics[width=8.2cm]{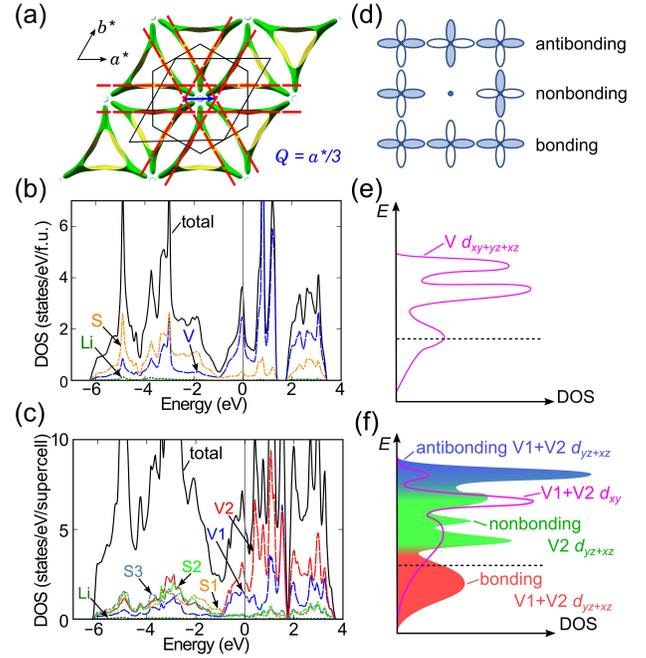}
\caption{\label{fig:Fig4}
(Color online) (a) Fermi surface in high-temperature phase of Li$_{0.33}$VS$_2$. Partial density of states in (b) high- and (c) low-temperature regions. (d) Three molecular orbitals derived accompanied by the 3c-2e formation. Schematic view of band structures originating from $d_{xy}$, $d_{yz}$ and $d_{zx}$ orbitals at (e) high and (f) low temperatures. }  
\end{figure}

A characteristic Fermi surface to bolster this argument is realized by a first principles calculation. As shown in Figure \ref{fig:Fig4}(a), the Fermi surface at high temperature facilitates a good nesting feature with $Q$ = $a$$^*$/3, which corresponds to the lattice periodicity in the low-temperature phase. Although Li-ions one-dimensionally aligned parallel to the ribbon chain leads us to suspect that Li-ion ordering is a possible driving force behind the phase transition at 375 K, this is not the case. To support this claim, we assume a 3 $\times$ 1 $\times$ 1 supercell, and structurally optimized the system using the Virtual Crystal Approximation (VCA) scheme, whereby Li-ions are uniformly aligned. The result indicates that the ribbon chain structure actually reduces the ground state energy and yields bond lengths near the experimental lengths: $A_L$ = 3.28, $B_L$ = 3.06, and $C_L$ = 3.73. This means that the instability towards the ribbon chain structure is inherent in the electronic state of the two-dimensional VS$_2$ plane and the observed Li-ion ordering is merely a consequence of the ribbon chain formation.

We should note that there are some tellulides,  $M$Te$_2$ ($M$ = V, Nb, Ta), which exhibit structural phase transition from a regular triangular lattice with trigonal symmetry to monoclinic with the formation of ribbon chains upon cooling \cite{rf:10,rf:19,rf:19-2}. It is discussed that the Te $p$-block bands increase in energy because of the interlayer Te-Te interactions, thereby leading to a partial electron transfer from the top portion of the Te $p$-block bands to the $d$-block bands of the metal, resulting in the actual $d^{4/3}$ electron count and following 3c-2e bond formation. However, the present Li$_{0.33}$VS$_2$ should be more appropriate than these tellulides for investigating the intrinsic nature derived from loose 3c-2e formation for some reasons. Initially, the formal $d$ electron count for vanadium can be finely tuned to 4/3 in Li$_{0.33}$VS$_2$. The diffraction study clearly shows that interlayer S-S distance of 3.788 \AA~at 400 K is significantly longer than the sum of van der Waals radii of S$^{2-}$ ion \cite{rf:24}, which means that the interlayer $p$-$p$ hybridization is negligibly small compared with these tellulides. In turn, this indicates that the $d$ electronic state can be finely tuned by controlling the Li fraction. The neutron diffraction result indicates that the Li fraction is quite close to 1/3, which ensures the presence of $d^{4/3}$ electronic state in Li$_{0.33}$VS$_2$. Secondary, we can expect weak $p$-$d$ hybridization in Li$_{0.33}$VS$_2$, which is compared with strong $p$-$d$ hybridization inherent to 1T tellurides \cite{rf:23}. The previous band calculations clearly indicate the considerable contribution of Te 5$p$ character at the Fermi level \cite{rf:21}, which prevents us from studying the intrinsic nature of 3c-2e state derived from strong $d$ character. In Li$_{0.33}$VS$_2$, the first principles calculation clearly indicates the negligibly small $p$-$d$ hybridization. As shown in Figs. \ref{fig:Fig4}(b) and \ref{fig:Fig4}(c), the S 3$p$ band is completely separated from the V 3$d$ band in both the high- and low-temperature phases, and the strong V 3$d$ character dominates near the Fermi energy. 

The calculation further clarifies the modification of the band structure derived from the 3c-2e formation. As shown schematically in Fig. \ref{fig:Fig4}(e), the $xy$, $yz$, and $zx$ bands are triply degenerated at high temperatures. Thus, we expect the $yz$ and $zx$ bands to transform into bands composed of three components at a low temperature: bonding, nonbonding, and antibonding, as shown in Fig. \ref{fig:Fig4}(d). We thus divide the $yz$ and $zx$ bands into three parts, as shown in Fig. \ref{fig:Fig4}(f). Whereas the bonding orbitals are almost filled, the Fermi surface survives due to the small stabilization energy of the bonding orbital following the incomplete gap opening between the bonding and nonbonding orbitals, and the overlap between the $yz$ and $zx$ bands, with the $xy$ band further generating a small charge transfer between them. The existence of the Fermi surface is consistent with the metallic conductivity observed in the resistivity measurement.

The metallic conductivity at low temperatures differs sharply from the insulating behavior observed in conventional molecular cluster compounds with 2c-2e bonds \cite{rf:2,rf:3,rf:4,rf:5,rf:6,rf:7,rf:8,rf:9}. The reason is strongly related to the bonding nature. For example, LiVS$_2$, which is the $d^2$ analog of Li$_{0.33}$VS$_2$, undergoes a metal-to-insulator transition at 314 K, accompanied by the formation of V triangular shaped trimers consisting of multiple 2c-2e bonds \cite{rf:4}. The change in V-V distance reaches $\sim$12\% at the transition in LiVS$_2$, whereas the V-V distance is 3.39\AA~at high temperatures and the intratrimer V-V distance decreases to 2.997\AA~at low temperatures \cite{rf:4}. The $\sim$12\% change in the V-V distance is much greater than the $\sim$8\%~change in Li$_{0.33}$VS$_2$, which indicates that the 2c-2e bond is much stronger than the 3c-2e bond, as can be understood intuitively. Because the bond strength is a function of the band gap, we consider that the weakness of the 3c-2e bond may be explained by the incomplete gap, leading to the metallic conductivity in the low-temperature phase of Li$_{0.33}$VS$_2$. Conversely, however, a large change in entropy of $\Delta$$S$ = 6.6 Jmol$^{-1}$K$^{-1}$ at the transition in Li$_{0.33}$VS$_2$ seldom appears in conventional itinerant electron systems, such as 2H-TaSe$_2$ and 4Hb-TaSe$_2$ \cite{rf:25}, which frequently appear in compounds in which molecular clusters form accompanied by orbital ordering \cite{rf:4,rf:26,rf:27,rf:28}. The coexistence of high electrical conductivity and a large change in entropy is a remarkable feature of Li$_{0.33}$VS$_2$ with 3c-2e bonds, which sharply differs from that for 2c-2e bonds, leading to new functional materials.

One possible application is as a heat-storage material with rapid thermal response. Many conventional heat-storage materials exploit a large change in enthalpy in the solid-liquid phase transition, such as paraffin with 140 Jcc$^{-1}$ at m.p. = 64 $^\circ$C \cite{rf:29} and polyethylene glycol with 165 Jcc$^{-1}$ at m.p. = 20 $^\circ$C \cite{rf:30}. Heat-storage materials that exploit a solid-liquid phase transition have a large enthalpy, whereas a low thermal conductivity leads to a difference between the internal temperature and the surface temperature, making it difficult to keep the surface temperature constant. Thus, fabricating heat-storage materials that exploit a solid-solid phase transition is strongly desired, and attention has focused on heat-storage materials that exploit the enthalpy change that accompanies by the electronic phase transition in strongly correlated electron systems \cite{rf:34, rf:31, rf:32, rf:33}. Our estimate gives an enthalpy change $\Delta$$H$ = 73.1 Jcc$^{-1}$ for Li$_{0.33}$VS$_2$, which is derived from $\Delta$$S$ = 6.6 Jmol$^{-1}$K$^{-1}$. This is a relatively large value for a heat-storage material that exploits a solid-solid transition, as summarized in Table SII in the supplementary material \cite{rf:Supple3}. Note that the high electrical conductivity, which is a feature of Li$_{0.33}$VS$_2$, improves the thermal conductivity and leads to a rapid thermal response. This characteristic is unique to Li$_{0.33}$VS$_2$ and is not found in conventional heat-storage materials that use the solid-solid phase transition. Thus, the 3c-2e bond is an interesting topic for both basic and applied science. 

To summarize, we present physical and structural studies of the layered two-dimensional triangular lattice system Li$_{0.33}$VS$_2$, and discuss how Li$_{0.33}$VS$_2$ embodies the 3c-2e bond. Compared with the conventional 2c-2e bond, the 3c-2e bond is loose, which leads to an incomplete band gap and the concomitant metallic conductivity at low temperature. In contrast, the entropy change at the transition is exceptionally large as a metal. The coexistence of these features may make Li$_{0.33}$VS$_2$ a novel and useful functional material.

\begin{acknowledgments}
The authors are grateful to Y. Okamoto and N. Mitsuishi for valuable discussions. This work was partly supported by a Grant in Aid for Scientiﬁc Research (Nos. 17K05530 and 17K17793), Fuji Science and Technology Foundation, The Thermal and Electric Energy Technology Inc. Foundation and Daiko Foundation. This work was carried out under the Visiting Researcher's Program of the Institute for Solid State Physics, the University of Tokyo. The synchrotron powder X-ray diffraction experiments were conducted at the BL5S2 of Aichi Synchrotron Radiation Center, Aichi Science and Technology Foundation, Aichi, Japan (Proposal No.201702049, 201702101, 201703027, 201704027 and 201704099). The neutron experiment at the Materials and Life Science Experimental Facility of the J-PARC was performed under a user program (Proposal No.2017B0041). Sample preparation for the neutron diffraction measurements was done at the CROSS user laboratories. 
\end{acknowledgments}

\end{document}